\documentclass[aps,pre,12pt,showpacs]{revtex4-1}
\usepackage{graphicx}
\usepackage{epstopdf}
\usepackage{amsmath}
\usepackage{amssymb}
\usepackage{default}
\usepackage{default}
\newcommand{\be}{\begin{equation}}
\newcommand{\bea}{\begin{eqnarray}}
\newcommand{\bc}{\begin{center}}            
\newcommand{\ee}{\end{equation}}
\newcommand{\eea}{\end{eqnarray}}
\newcommand{\ec}{\end{center}}

\newcommand{\baa}{\begin{eqnarray*}}
\newcommand{\eaa}{\end{eqnarray*}}
\begin{document}

\title{Efficiencies of power plants, quasi-static models
and \\ the geometric-mean temperature}

\author{Ramandeep S. Johal}
\email{rsjohal@iisermohali.ac.in}
\affiliation{ Department of Physical Sciences, \\ 
Indian Institute of Science Education and Research Mohali,\\  
Sector 81, S.A.S. Nagar, Manauli PO 140306, Punjab, India}
\begin{abstract}
Observed efficiencies of industrial power plants are often
approximated by the square-root formula: $1-\sqrt{T_-/T_+}$,
where $T_+ (T_-)$ are the highest (lowest) temperature
achieved in the plant.
This expression can be derived within finite-time thermodynamics,
or, by entropy generation minimization, based on finite rates
of processes. A closely related quantity is the optimal 
value of the intermediate temperature for the hot stream,
which is given by the geometric-mean value: $\sqrt{T_+ T_-}$. 
It is proposed to model the operation of plants by 
quasi-static work extraction models, with one reservoir
(source/sink) as finite, while the other as practically
infinite. No simplifying assumption is made on the nature 
of the finite system. This description is consistent with two model
hypotheses, each yielding a specific value of the intermediate
temperature. We show that the expected value of the 
intermediate temperature, defined as the arithmetic mean,
is very closely given by the geometric-mean value.
The definition is motivated as the use of inductive  
inference in the presence of limited information.
\end{abstract}

\maketitle
{\bf Introduction:}
In recent years, there has been a great interest in extending  
thermodynamic models to justify the observed performance of industrial
power plants \cite{Curzon1975, Bejan1997, Esposito2010, Moreau2012}.
The observed efficiencies are usually much less than 
the Carnot limit 
\be
\eta_C = 1- \frac{T_-}{T_+},
\label{ec}
\ee
evaluated on the basis of the highest ($T_{+}$)  and the 
lowest ($T_{-}$) temperatures, for the particular plant.
Naturally, real machines operate under irreversibilities
caused by various factors, like finite rates of heat transfer and fluid flow, 
internal friction, heat leakage and so on, 
unlike the idealized quasi-static processes of a reversible cycle.
Thus the analysis of irreversible models with finite-rate
processes seems a reasonable goal to pursue.
 One often-studied measure is the efficiency at maximum
power of an irreversible model which is then compared with the observed
efficiency of these plants. 

The earliest known such model is ascribed to Reitlinger \cite{Reitlinger},
which involved a heat exchanger receiving heat from a finite hot stream
fed by a combustion process. An analogous model was applied to  a steam 
turbine by Chambadal \cite{Chambadal}. The considered heat exchanger 
in these models is effectively infinite. Novikov \cite{Novikov} considered 
the heat transfer process 
between a hot stream and a finite heat exchanger with a given heat conductance. 
Two, simple but significant, assumptions enter these models: 
i) constant specific heat of the inlet hot stream and ii) validity
of Newton's law for heat transfer. 
Further, there appears a floating, temperature variable  
in between the highest and the lowest values, such as the 
temperature of the exhaust warm stream, over which the output
power can be optimized. This yields an optimal value 
of the intermediate temperature which 
is usually found to be $\sqrt{T_+ T_-}$, the geometric mean of $T_+$ and $T_-$.
Related to this fact, is the conclusion that the 
efficiency at maximum power is given by an elegant expression:
\be
\eta_{CA} = 1-\sqrt{\frac{T_-}{T_+}}.
\label{eca}
\ee
Due to historical imperative \cite{Feidt2014}, the above expression may be called 
Reitlinger-Chambadal-Novikov efficiency. However,  more recently 
it was rediscovered by Curzon and Ahlborn (CA) \cite{Curzon1975}.
Thus in the physics literature,
it is more popularly addressed as CA-value. This latter model  
considered finite rates of heat transfer at both the hot and the cold contacts, 
but also explicitly considered the  times allocated to these contacts.
The average power per cycle may be optimized over these times \cite{Curzon1975},
or alternately, over the intermediate temperature variables \cite{Joubook}. This model  
spawned much activity and the new area borne thereforth was 
termed Finite-time Thermodynamics \cite{Andresen2011}.
In the engineering literature,
the corresponding approach is called Entropy Generation Minimization \cite{Bejan1996, Bejan1997}.

A positive indication 
for the simple thermodynamic approach is that the 
actually observed efficiencies of industrial plants happen to be quite close to CA-value. 
Fig.1 shows this comparison as tabulated
in Table I. Although the agreement is close, the observed values can be higher, or lower,
than CA-value. This apparent discrepency has stimulated further extensions of models,
for instance using the low-dissipation assumption \cite{Esposito2010},
 which predict the efficiency at maximum power, to be bounded as:
\be
\frac{\eta_C}{2} \le {{\eta}} \le \frac{\eta_C}{2 - \eta_C}.
\label{bnds}
\ee
\begin{figure}[ht]
\includegraphics[width =9cm]{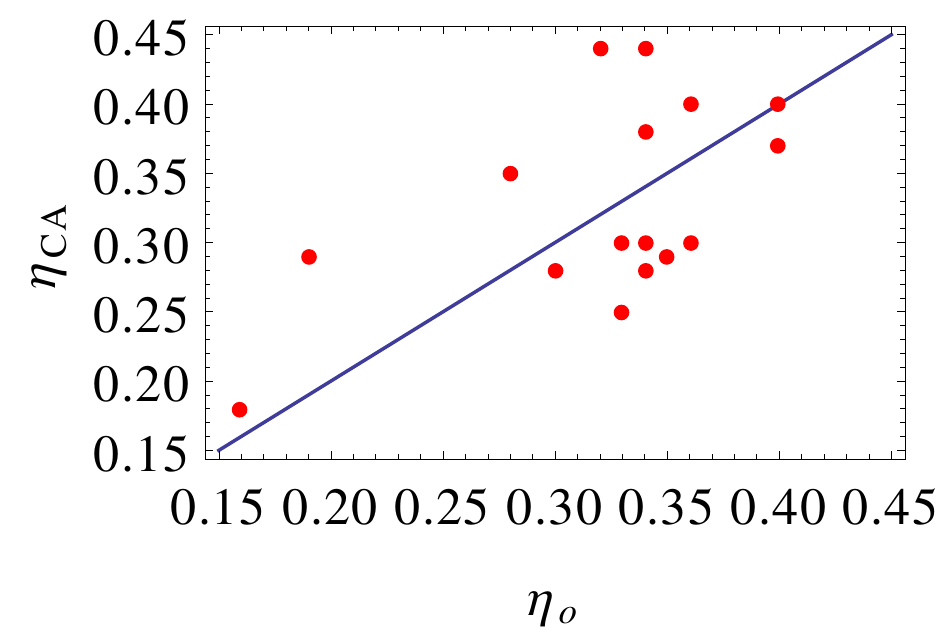}
 \caption{Data on the observed efficiencies ($\eta_o$) of power plants, plotted against the 
 CA-value, $\eta_{CA} = 1-\sqrt{T_-/T_+}$ for respective plants as given in Table I.
 A point lying on the straight line 
 has the observed efficiency equal to CA-value. 
 So for points above the line, $\eta_o$ is below
 the CA-value, while the converse is true for the points below this line.}
\label{ecaf}
 \end{figure}
It is then realized that most of the observed efficiencies fall within 
these bounds \cite{Esposito2010, Moreau2012}.
 Naturally, the question of 
the actual working constraints
and the real optimization targets for each plant, is also relevant. 
Still, the effectiveness of these simple models, in reproducing the gross
features of diverse plants, cannot be denied. 
\begin{table}[ht]
\begin{tabular}{lccccccccc}
\hline
Industrial Plant &\quad $T_+$ \quad &\quad $T_-$\quad & \quad $\eta_o$ \quad & \quad $\eta_{CA}$ 
\quad & \quad  $T_{m}^{(-)}$ \quad & \quad  $T_{m}^{(+)}$ \quad & \quad  $T_{m}^{(av)}$ 
\quad & \quad  $G(T_+,T_-)$   \\

 \hline \hline 
Almaraz II (Nuclear, Spain) \cite{Bejan1997}\quad & \quad  600 \quad & \quad  290 
\quad & \quad .34 \quad & \quad .30
\quad & \quad 439.39 \quad & \quad  396.0 \quad & \quad  417.7 \quad & \quad  417.13  \\
Calder Hall (Nuclear, UK) \cite{Bejan1997}\quad & \quad  583 \quad & \quad  298 
\quad & \quad .19 \quad & \quad .29
\quad & \quad 367.9  \quad & \quad  472.23\quad & \quad  420.07\quad & \quad  416.81   \\
CANDU (Nuclear, Canada) \cite{Curzon1975} \quad & \quad  573 \quad & \quad  298 
\quad & \quad .30 \quad & \quad .28
\quad & \quad 425.71 \quad & \quad  401.1 \quad & \quad  413.41\quad & \quad  413.22 \\
Cofrentes (Nuclear, Spain) \cite{Bejan1997} \quad & \quad  562 \quad & \quad  289 
\quad & \quad .34 \quad & \quad .28
\quad & \quad 437.88 \quad & \quad  370.92\quad & \quad  404.40\quad & \quad  403.01 \\
Doel 4 (Nuclear, Belgium) \cite{Bejan1997}\quad & \quad  566 \quad & \quad  283 
\quad & \quad .35 \quad & \quad .29
\quad & \quad 435.39 \quad & \quad  367.9 \quad & \quad  401.64\quad & \quad  400.22 \\
Heysham (Nuclear, UK) \cite{Bejan1997} \quad & \quad  727 \quad & \quad  288 
\quad & \quad .40 \quad & \quad .37
\quad & \quad 480.0  \quad & \quad  436.2 \quad & \quad  458.1 \quad & \quad  457.58 \\
Larderello (Geothermal, Italy) \cite{Curzon1975} \quad & \quad 523\quad & \quad 353 
\quad & \quad .16 \quad & \quad .18
\quad & \quad 420.24 \quad & \quad  439.32\quad & \quad  429.78\quad & \quad  429.67 \\
Sizewell B (Nuclear, UK) \cite{Bejan1997}   \quad & \quad  581 \quad & \quad  288 
\quad & \quad .36 \quad & \quad .30
\quad & \quad 450.0  \quad & \quad  371.84\quad & \quad  410.92\quad & \quad  409.06 \\
West Thurrock (Coal, UK) \cite{Curzon1975} \quad & \quad  838 \quad & \quad  298 
\quad & \quad .36 \quad & \quad .40
\quad & \quad 465.63 \quad & \quad  536.32\quad & \quad  500.97\quad & \quad  499.72 \\
Pressurized water nuclear reactor \cite{Borlein2009} \quad & \quad  613 \quad & \quad  304
\quad & \quad .33 \quad & \quad .30 
\quad & \quad 453.73 \quad & \quad  410.71\quad & \quad  432.22\quad & \quad  431.69 \\
Boiling water nuclear reactor \cite{Borlein2009} \quad & \quad  553 \quad & \quad  304
\quad & \quad .33 \quad & \quad .25
\quad & \quad 453.73 \quad & \quad  370.51\quad & \quad  412.12\quad & \quad  410.02 \\
Fast neutron nuclear reactor \cite{Borlein2009}\quad & \quad  823 \quad & \quad  296 
\quad & \quad .40 \quad & \quad .40
\quad & \quad 493.33 \quad & \quad  493.8 \quad & \quad  493.57\quad & \quad  493.57 \\
(Steam/Mercury, US) \cite{Bejan1997} \quad & \quad  783 \quad & \quad  298 
\quad & \quad .34 \quad & \quad .38
\quad & \quad 451.52 \quad & \quad  516.78\quad & \quad  484.15\quad & \quad  483.05 \\
(Steam, UK) \cite{Bejan1997} \quad & \quad  698 \quad & \quad  298 
\quad & \quad   .28 \quad & \quad .35
\quad & \quad 413.89 \quad & \quad  502.56\quad & \quad  458.22\quad & \quad  456.08 \\
(Gas Turbine, Switzerland) \cite{Bejan1997} \quad & \quad  963 \quad & \quad  298
\quad & \quad .32 \quad & \quad .44
\quad & \quad 438.24 \quad & \quad  654.84\quad & \quad  546.54\quad & \quad  535.7 \\
(Gas Turbine, France) \cite{Bejan1997} \quad & \quad  953 \quad & \quad  298 
\quad & \quad .34 \quad & \quad .44
\quad & \quad 451.52 \quad & \quad  628.98\quad & \quad  540.25 \quad & \quad  532.91\\
\hline
\end{tabular}
\caption{Observed efficiencies ($\eta_o$) of power plants working between
temperatures $T_+$ and $T_-$, compared with $\eta_{CA} = 1 -\sqrt{T_-/T_+}$.
The effective temperatures are defined as:
$T_{m}^{(+)} = T_+ (1-\eta_0)$, $T_{m}^{(-)} = T_-/(1-\eta_0)$, 
$T_{m}^{(av)} = (T_{m}^{(+)} + T_{m}^{(-)} )/2$, and $G(T_+, T_-) = \sqrt{T_+ T_-}$.}  
\end{table}

Apart from finite-time models, the irreversibilities
reducing the efficiency to a lower-than-Carnot value, may also
be treated within a quasi-static work extraction models 
with finite source/sink \cite{Thomson, Ondrechen1981, Leff1986, Lavenda2007}.
Some of these studies also derive the optimal value of an intermediate temperature
which is the geometric-mean value, and consequently the 
efficiency at maximum work equals the CA-value.
However, here again, the simplifying assumption 
of a constant heat capacity, say, of the source
or the working medium, heavily determines the conclusion.

More recently, it was observed \cite{JohalRai2016} that within 
linear response theory, the 
bounds such as Eq. (\ref{bnds}), also follow
within a quasi-static model of work extraction from  
finite-sized heat source and sink. Ref. \cite{JohalRai2016} makes no specific assumption
about the nature of the heat source/sink. 
The present study is motivated by this analysis, and aims to study 
the performance of real plants within a quasi-static approach.
More precisely, we address an inverse question. Instead of 
finding the optimal intermediate temperature and then 
from it, the efficiency of the process, we use the information
on the highest and the lowest temperatures, along with the value of
observed efficiency, to infer the intermediate temperature
at maximum work. The correspondence between this temperature and 
efficiency is as follows. If the former is exactly equal to
the geometric-mean value, then the efficiency is equal to
CA-value.
The model of work extraction is based on 
one system as a finite source, or sink, and the other as an infinite reservoir,
which allows
for two alternate scenarios: i) when the source is finite and ii) when
the sink is finite. The limited information on the actual situation
being realized out of  these two possibilities, motivates to do an inference analysis.
Thus we estimate an expected value of the intermediate temperature
scale. Interestingly, this value is found to be very close to the geometric
mean of $T_+$ and $T_-$. We also present a quantitative
argument for this proximity to the geometric-mean value. Thus our analysis 
indicates the role of geometric mean from a  novel angle
which has two distinctive features: first, it is based on maximum work 
approach, and second, we do not per se assume a specfic nature of 
the finite source or sink. 

Our starting point
is the reversible cycle operating between two infinite heat 
reservoirs, for which the efficiency is the Carnot value 
$\eta_C$. As a first step marking a deviation from reversibility, 
we consider one reservoir as finite, while the other reservoir remains very large
compared to the former, or practically infinite.
\begin{figure}[ht]
\includegraphics[width =9cm]{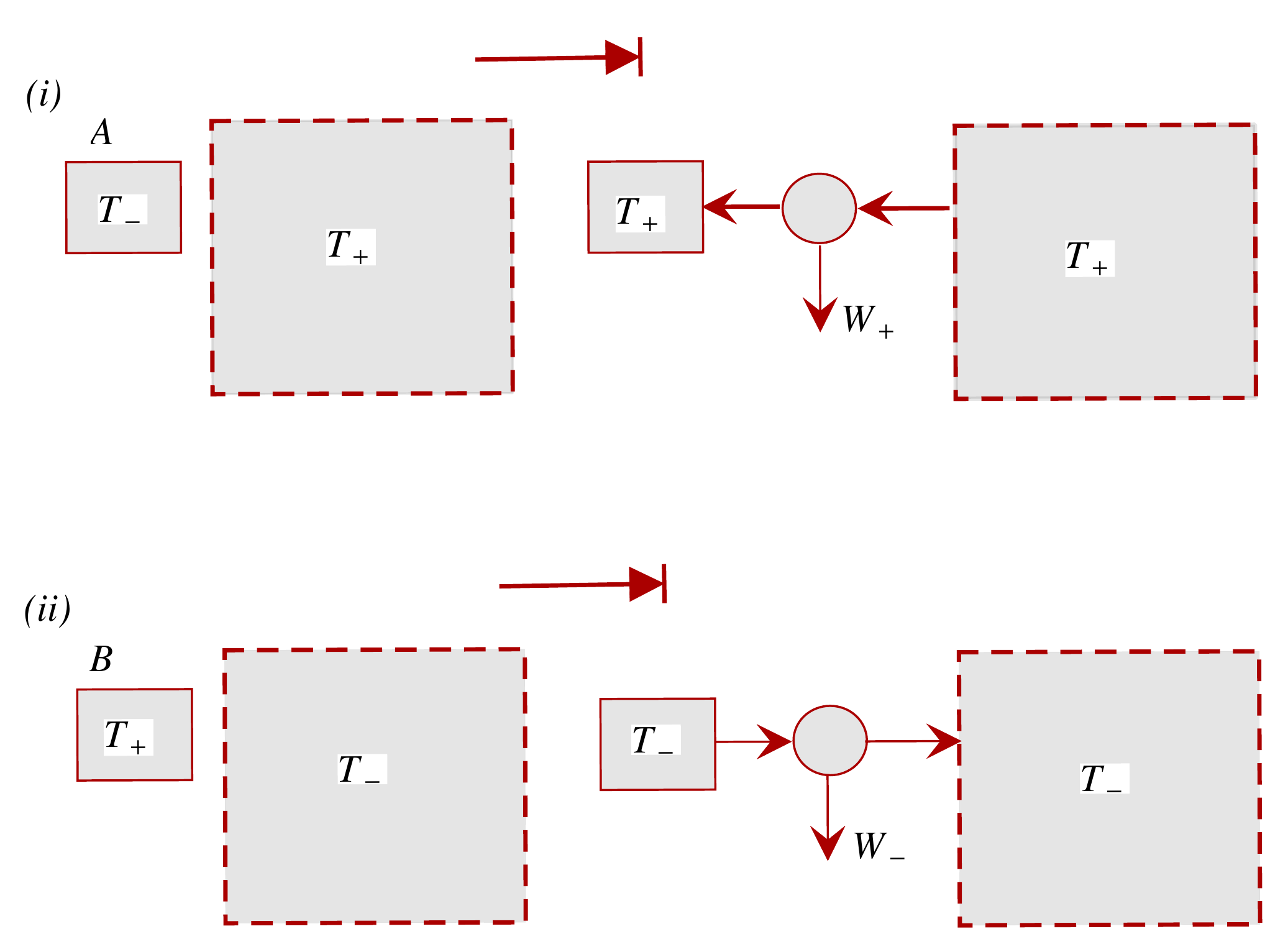}
 \caption{Schematic of the engine between a finite system and a heat reservoir, 
  for a given pair of initial temperatures $(T_+,T_-)$: 
  (i) System A as a finite sink at $T_-$ and an infinite source at $T_+$,
  coupled with a reversible work source. Work extraction $W_+$, Eq. (\ref{wp}),
  is completed when the temperature
  of A becomes $T_+$. (ii) System B as a finite source at $T_+$ and an infinite sink
  at $T_-$. Maximum
  extracted work is $W_-$, Eq. (\ref{wm}), when the temperature of B becomes $T_-$.}
\label{engine}
 \end{figure}
Now, we first assume that system A acts as a finite heat sink at temperature $T_-$,
relative to a very large heat source at temperature $T_+$.
The two are coupled by an ideal engine which delivers work to 
a reversible work source, by running in 
infinitesimal heat cycles, that successively increase
the temperature of A, till A comes in
equilibrium with the source, see Fig.1 (i). At this point,
the extracted work is maximum. Suppose that in this process,
the system A moves from an equilibrium state characterized by
energy $U_-$ and entropy $S_-$ to an equilibrium state with the
corresponding values of $U_+$ and $S_+$.
The heat removed from the source is $Q_+ =   T_+ (S_+ - S_-)$.
The heat rejected to the finite sink is $q_+ = U_+ - U_-$. 
The work extracted is given by $W_+ = Q_+ - q_+$, or 
\be
W_+ =  T_+ (S_+ - S_-) - (U_+ - U_-).
\label{wp}
\ee
Then the efficiency at maximum work, $\eta_+ = W_+ /Q_+$,
is evaluated to be:
\be
\eta_+ = 1- \frac{T_{m}^{(+)}}{T_+},
\label{ep}
\ee
where we have defined 
\be
 T_{m}^{(+)} = \frac{U_+ - U_-}{S_+ - S_-},
\label{tm}
 \ee
 a quantity characteristic of system A.
Further $T_{m}^{(+)}$ may be regarded 
as an effective temperature of an infinite reservoir \cite{Johal2016},
such that the present situation of an infinite source and a finite sink, 
at temperatures $T_+$ and $T_-$ respectively, is equivalent 
to maximum work extraction in a reversible cycle from two infinite reservoirs
at temperatures $T_+$ and $T_{m}^{(+)} (< T_+)$. In the latter case, the extracted work per 
 cycle is: $W_+ = (T_+ - T_{m}^{(+)}) (S_+ - S_-)$, with (Carnot) efficiency 
$1-T_{m}^{(+)}/T_+$, which is the same as Eqs. (\ref{wp}) and (\ref{ep}).

\begin{figure}
\includegraphics[width =9cm]{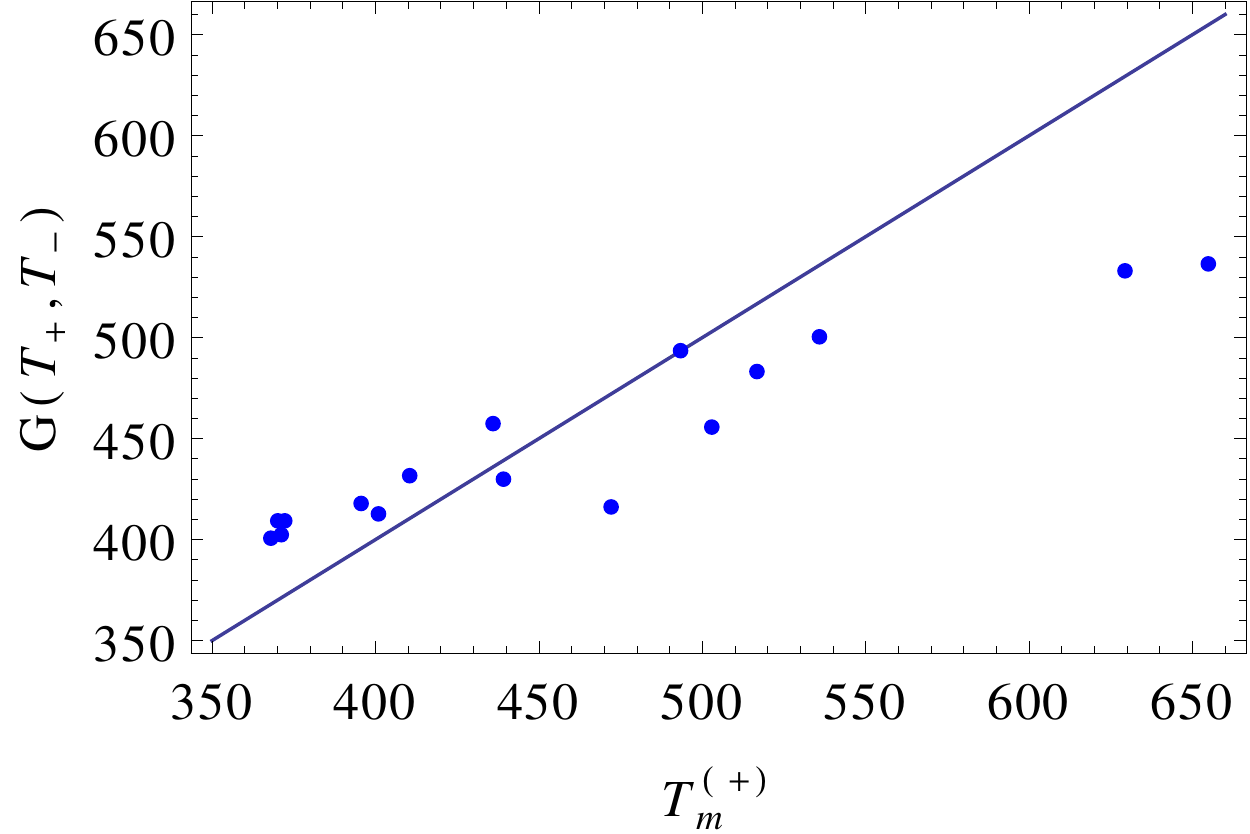}
 \caption{The effective temperature $T_{m}^{(+)}$ plotted against
 the geometric mean $G(T_+,T_-) = \sqrt{T_+ T_-}$, the two quantities being
 equal along the straight line.}
\label{tmpt}
 \end{figure}

Now, we assume that the information on the nature of  system A
is not available, or, in other words, $T_{m}^{(+)}$ is not given.
But if  the quasi-static model with an infinite source
and a finite sink, is a good model for the observed performance 
of an industrial plant, then 
we may infer the relevant value of the effective
temperature $T_{m}^{(+)}$, when the theoretical   
efficiency ($\eta_+$) for the model is set equal to the observed
value $\eta_o$. This implies that we write 
\be
T_{m}^{(+)} = T_+ (1-\eta_o).
\label{tmpo}
\ee
Such values for some industrial plants, based  
on the observed values of their efficiencies, 
are tabulated in Table I, and also depicted in Fig. \ref{tmpt}, 
in comparison to geometric mean $G(T_+, T_-) = \sqrt{T_+ T_-}$. 
The latter value is chosen simply because
when $T_{m}^{(+)} = G$, the 
observed efficiency is equal to the CA-value. Thus the spread of observed
efficiency values around the CA-value in Fig. \ref{ecaf},  
is translated here, to a spread in the effective temperatures
around the geometric-mean values. More precisely, $T_{m}^{(+)} \gtrless G$
indicates $\eta_o \lessgtr \eta_{CA}$, as is evident from Table I.
It is to be noted that inferring the effective temperature
from the observed efficiency does not determine the nature
of the system A or the form of fundamental relation $U(S)$.
The latter information is not at our disposal within the 
assumptions of the model.  

\begin{figure}[ht]
\includegraphics[width =9cm]{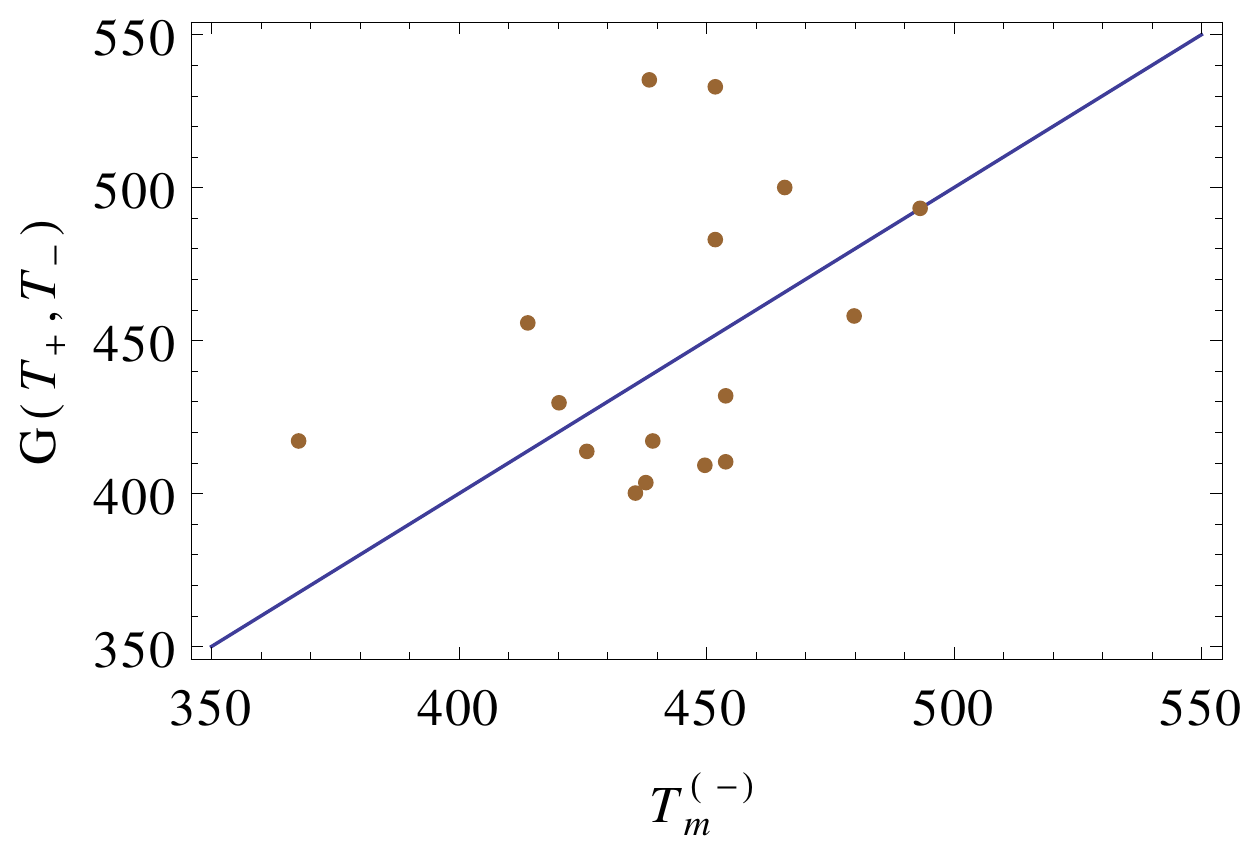}
 \caption{The effective temperature $T_{m}^{(-)}$ plotted against
 the geometric mean $G(T_+,T_-) = \sqrt{T_+ T_-}$, the two quantities being
 equal along the straight line.}
\label{tmmf}
 \end{figure}
However, the scenario of work extraction from a finite system coupled to an 
infinite reservoir, via a reversible process, is consistent with 
an alternate picture too.
This involves that, for the same initial temperatures $T_+$
and $T_-$, we consider a finite source (B) at $T_+$,
coupled to an infinite sink at $T_-$, see Fig. \ref{engine} (ii).
Here again, we can extract maximal work by utilizing the temperature 
gradient between B and the reservoir, till B  comes to be
at temperature $T_-$ \cite{Izumida2014}. 
Assuming that the system goes from some equilibrium state $(U_+',S_+')$
to another one $(U_-',S_-')$, the heat removed from the source is $Q_- =  U_+' - U_-'$
and the amount rejected to sink is $q_- = T_- (S_+' - S_-')$. So
the extracted work is $W_- = Q_-' - q_-'$, or 
\be
W_- = (U_+' - U_-') - T_- (S_+' - S_-').
\label{wm}
\ee
This is called {\it exergy} in the engineering literature \cite{Exergy}.
The efficiency $\eta_- = W_- /Q_-$ is given by
\be
\eta_- = 1- \frac{T_-}{T_{m}^{(-)}},
\label{em}
\ee
where
\be
 T_{m}^{(-)} = \frac{U_+' - U_-'}{S_+' - S_-'}.
\label{tmp}
 \ee
It is clear from the expressions for $W_-$ and $\eta_-$,
that an equivalence exists between the above model and that of work extraction in 
a reversible cycle from two infinite reservoirs
at $T_-$ and $T_{m}^{(-)} (> T_-)$. 

Now again, to apply the above case to model an industrial plant, we can equate 
the observed efficiency to the theoretical efficiency, as $\eta_o =\eta_-$
and infer the corresponding effective temperature $T_{m}^{(-)}$
from Eq. (\ref{em}), as 
\be
T_{m}^{(-)} = \frac{T_-}{1-\eta_o}.
\label{tmm}
\ee
It is clear that $T_+ > T_{m}^{(-)} > T_-$.
The calculated values of $T_{m}^{(-)}$, based on the observed efficiencies
of the plants, are also tabulated in Table I, and shown graphically
in Fig. \ref{tmmf} in comparison to $G(T_+, T_-)$.  

The following remark seems appropriate here. An analogy 
may be made between the above model and the earlier irreversible models such as proposed
by Chambadal \cite{Chambadal, Bejan1997},
in which an intermediate temperature $T_{w}$ (of the warm
exit stream) enters the analysis and the theoretical efficiency of the model 
is $1-T_-/T_w$ (compare with Eq. (\ref{em})). For an optimal value  $T_w =\sqrt{T_+ T_-}$, 
the power output becomes optimal with the corresponding efficiency equal to CA-value.
The crucial differences, from our model, are twofold:
a) we study quasi-static maximum work extraction process b) the nature of
system B, the finite source, is not specified. 

\begin{figure}[ht]
\includegraphics[width =10cm]{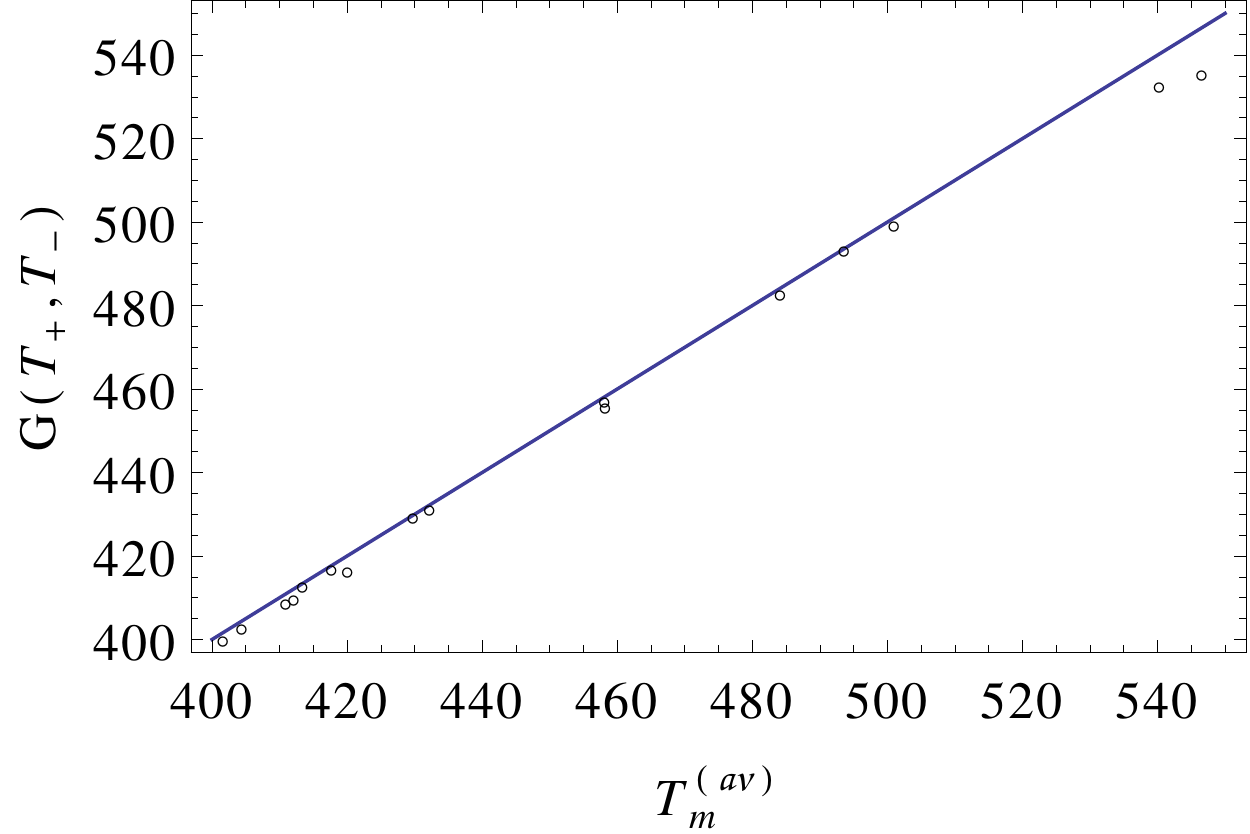}
 \caption{The average effective temperature $T_{m}^{(av)}$ plotted against
 the geometric mean $G(T_+,T_-) = \sqrt{T_+ T_-}$, the two quantities being
 equal along the straight line. $T_{m}^{(av)}$ values calculated from the observed data on
 most of the power plants, is remarkably close to $G(T_+,T_-)$.}
\label{gmtm}
 \end{figure}

Now as observed in Figs. \ref{tmpt} and \ref{tmmf}, the values of
effective temperatures seem to be distributed, apparently in a
random fashion, about the geometric-mean value. However, it 
is remarkable to note that for a given plant, the calculated
values of $T_{m}^{(+)}$ and $T_{m}^{(-)}$ are such that, to a high accuracy, they are 
equidistant from the geometric mean $G$.
More precisely, if we define an average value of temperature
$T_{m}^{(av)}$, as $T_{m}^{(+)} - T_{m}^{(av)} = T_{m}^{(av)} - T_{m}^{(-)}$,
then this value $T_{m}^{(av)}$ is very close to the
$G$ value for that situation.  
In other words, we define an average scale of temperature
as the arithmetic mean of 
 the two inferred temperatures $T_{m}^{(\pm)}$, and so given by
\be 
T_{m}^{(av)} = \frac{1}{2}\left[ T_+ (1-\eta_o)  +   \frac{T_-}{1-\eta_o} \right].  
\label{tmav}
\ee
Then the above average value 
is found to be closely approximated
by $G(T_+,T_-)$ for most observed cases of industrial plants,
see Fig. \ref{gmtm}, as well as Table I.

Now, we turn to a more quantitative characterization of the
above observation.  
It is easy to see that $T_{m}^{(av)}$ takes a minimum value of  $\sqrt{T_+ T_-}$,
w.r.t to $\eta_o$ ($d T_{m}^{(av)}/d \eta_o =0$, and $d^2 T_{m}^{(av)}/d \eta_o^2 >0$),
at $\eta_o = 1- \sqrt{T_-/T_+}$. 
This implies that any possible value of $T_{m}^{(av)}$ is equal to or greater than
$\sqrt{T_+ T_-}$, and so will lie on or below 
the straight line plotted in Fig. \ref{gmtm}. 

The second aspect is related to the observation made earlier 
that the deviations in values of effective temperatures $T_{m}^{(\pm)}$
from the corresponding $G$-values, reflect the fact 
how the observed efficiency deviates from the 
CA value. However, the deviations from $G$-values, are suppressed considerably in
 case of $T_{m}^{(av)}$. 
This may be argued by considering small deviations ($\epsilon$) 
in efficiency from the CA value, and expanding $T_{m}^{(av)}$
in powers of $\epsilon$. Let $\eta_o = \eta_{CA} + \epsilon$.
Then we see that up to second order:
\be 
T_{m}^{(av)} = \sqrt{T_+ T_-} + \frac{1}{2}\sqrt{\frac{T_{+}^{3}}{T_{-}}} 
\epsilon^2 + {\cal O}(\epsilon^3).
\ee
Thus, the first non-zero correction from the geometric-mean value is of second 
order in $\epsilon$, while it is of the first order for 
$\eta_o$, or $T_{m}^{(\pm)}$. Clearly, the magnitude of fluctuations is suppressed
in the case of $T_{m}^{(av)}$.

Finally, we address the meaning of the average effective temperature.
If we again consider the two extreme situations envisaged in Fig. \ref{engine},
then they are mutually exclusive, or one may say, they are counterfactual.
The average temperature is not necessarily seen in an actual realization,
except for the special case $T_{m}^{(+)} = T_{m}^{(-)} = \sqrt{T_+ T_-}$.
In this sense, the meaning one can attach to the definition of 
$T_{m}^{(av)}$, can be given best in the language of inductive inference \cite{Jeffreys1939, Jaynes2003}.  
In latter terms, the average temperature 
represents an expected scale of the effective temperature, in view of our
inability to choose between the two alternatives (i) and (ii),
where each represent our hypothesis for the model of work extraction 
applicable to the plant.
In inference, when one of, say, two mutually exclusive hypotheses
cannot be given a preference over the other, then we must assign equal weights
to the inferences derived from each \cite{Laplace1774, Jaynes2003}. The effective temperatures
$T_{m}^{(\pm)}$ are our inferred values from the given data on the observed
efficiency, and, so $T_{m}^{(av)}$, defined with equal weights assigned  
to both the inferences, represents the expected
value of the effective temperature, commensurate with the information
at our disposal. Any deviation from equal weights would imply
that we have some extra piece of information about the process,
and thus would be inconsistent \cite{Jaynes2law}.

Concluding, we have proposed to model the observed efficiencies,
using the quasi-static models of work extraction where one 
reservoir (hot or cold) is finite while the other is practically infinite.
This brings an extra 
scale of intermediate temperature into the analysis.
We note two consistent models depending on which 
reservoir is taken as finite, and consequently,  
we get two possible values of the intermediate temperature. 
The limited information on the working conditions
does not allow to prefer one model hypothesis
over the other and so an equally-weighted average 
represents the expected scale of intermediate temperature.
For most of the power plants, this inferred
value is found to be quite close to the geometric
mean of the highest and the lowest temperatures. 
Thus we rediscover the significance of the geometric-mean
temperature, which was emphasized in earlier
irreversible models of plants \cite{Reitlinger,Chambadal,Novikov, Yvon}, 
but there it was often based on simplifying assumptions such as constant heat capacity
for the hot stream, and Newton's law for heat transfer.
Such assumptions are not important in our analysis
and we do not make use of a particular nature of
the finite reservoir. In the present approach, if the deviations in
the observed efficiency from the CA-value are small, then 
the geometric-mean temperature appears as a rational estimate
for the intermediate temperature.
  
An inference based approach has been used earlier 
 to study models of thermodynamic machines with limited information.
 The emergence of CA efficiency from inference  
 has been noted in quantum
 heat engine \cite{Johal2010} and mesoscopic models like Feynman's ratchet \cite{George2015}. 
 For classical models of work extraction
 from two finite reservoirs, the results for efficiency at maximum
 work are reproduced through inference, beyond linear
 response \cite{Aneja2013}.
 Further, reversible models with limited information have been related
 to irreversible models through inference based reasoning \cite{JRM2015}.
In the present context, it is remarkable how simple, but general 
considerations can lead to estimates close to the geometric-mean value 
for the intermediate temperature. 
 It is indeed curious to ask,
how the inference approach leads to the same conclusions
as the simple irreversible models with constant heat capacity. 
Does it indicate a connection between the  
simplicity of objective modelling and assuming minimal information
in inference?  More research is needed for a deeper understanding of  
 the connection between the use of limited information and 
 thermodynamic modelling.
 
The purpose of inductive inference is not just to 
provide an academic and philosophical point of view 
 which may sometimes rival 
 the more objective modelling we usually resort to
 in science,
 but also seek how methods of inference 
 can give insight into the actual state of affairs,
 while incorporating the prior information, normally 
 not considered in thermodynamic models. 
 In this context, we note that
 although most of the data on plants yield an 
 expected value of the intermediate temperature
 close to geometric-mean value, still there are 
 a couple of not-so-insignificant deviations in Fig. \ref{gmtm}, near the top
 right corner. A valid question is, why do these
 examples differ from the rest of the cases? Does it indicate
 other measures of optimization being
 used in the actual operation? As far as the 
 available data is concerned, we note  
 these plants operate under higher temperature
 gradients than most other plants. Thus, 
 one may argue that our model and the assumptions
 may not be good approximations for large 
 temperature differences. In any case, once
 we have noticed deviations from the expected
 behavior in these plants, further studies on the
 actual working conditions may
 be undertaken, which may then yield 
 information to generalize the inference analysis, or
 may help to improve the performance, at par with other plants. 

\end{document}